\documentclass[a4paper,11pt]{article}
\pdfoutput=1 
\newcommand{\comm}[1]{}
\usepackage{xcolor}
\usepackage{jinstpub} 
\usepackage{caption}
\usepackage{siunitx}
\usepackage{enumitem}
\usepackage{subcaption}

\notoc 

\title{\boldmath Optical Acceptance Testing of Serenity-S1 Boards for the CMS HGCAL Phase-2 Backend Electronics}

\collaboration[c]{on behalf of the CMS collaboration}

\author[a,1]{A. Qamesh,\note{Corresponding author.}} 
\author[b]{P. D. Dauncey,}
\author[a]{H. A. Krause,} 
\author[a]{P. Keskiner,}
\author[b]{S. Mallios,}
\author[b]{R. Shukla,}
\author[c]{N. Tarannum}
\affiliation[a]{Karlsruhe Institute of Technology, Institute for Data 
    Processing and Electronics,\\
    Hermann-von-Helmholtz-Platz 1, D-76344 
    Eggenstein-Leopoldshafen, Germany}
\affiliation[b]{Imperial College London, Physics Department, 
    Blackett Laboratory,\\
    Prince Consort Rd, London, SW7 2BW, U.K.}
\affiliation[c]{University of Bristol, School of Physics,
HH Wills Laboratory,\\
Tyndall Avenue
Bristol, BS8 1TL}

\emailAdd{ahmed.qamesh@cern.ch}

\abstract{The CMS Phase-2 upgrade includes a new high-granularity calorimeter. The calorimeter electronics will include around 300 Serenity-S1 FPGA boards that use Samtec FireFly electro-optical modules with a link speed of up to 25 Gbps. General production tests performed by the central Serenity group only validate the high-speed links via copper loopback cables. We present an optical acceptance test framework specifically designed for the calorimeter, providing comprehensive link and slow-control validation of installed FireFly modules under realistic operating conditions. This test combines loopback and board-to-board testing, extended bit error ratio measurements, and continuous module health monitoring within an automated workflow, delivering structured diagnostics and ensuring reliable large-scale system deployment. 

}


\proceeding{Topical Workshop on Electronics for Particle Physics\\
September 28 - October 2 2026\\
Castelldefels, Barcelona, Spain}

\begin{document}
\maketitle
\flushbottom
\section{Introduction}
\label{sec:intro}
The CMS High-Granularity Calorimeter (HGCAL) is one of the major
sub-detector upgrades for the High-Luminosity LHC~\cite{cms_phase2_daq}.
Its back-end electronics are built around the Serenity-S1
card~\cite{Mehner_2025}, an AMD VU13P FPGA platform with up to 124
optical links via Samtec FireFly modules~\cite{Mehner_2024}, shared across seven CMS
sub-detector systems. Of the more than 700 cards produced, 302 are allocated to HGCAL in four variants~\cite{cms_phase2_tdr}: \textbf{DAQ} (96 units, lpGBT at 10\,Gbps and SlinkRocket at 25\,Gbps)~\cite{cms_phase2_tdr};
\textbf{NZS} (2 units, CMS Standard Protocol (CSP) at 25\,Gbps)~\cite{cms_phase2_cms_trigger};
\textbf{Trigger Stage-1} (96 units, lpGBT RX at 10\,Gbps and CSP16
TX at 16\,Gbps); and
\textbf{Trigger Stage-2} (108 units, CSP16 at 16\,Gbps and
SlinkRocket at 25\,Gbps). 

The Serenity quality assurance chain starts with a Factory Acceptance Test
(FAT)~\cite{Krause_2026} after PCB assembly, repeated as part of a more comprehensive User
Acceptance Test (UAT) on delivery to CERN. Both use copper loopback
cables, because the FireFly modules are fitted later by the
sub-detector groups. Consequently, neither test validates optical-link operation, I$^2$C communication with the FireFly modules, or the integrity of their supply rails. 

The HGCAL Optical Acceptance Test (OAT) closes this
gap. It is the first test of the fully assembled boards with their
production FireFly modules in the final HGCAL configuration, and it
is a functional acceptance test of the assembly rather than a
component qualification.

\section{OAT Concept}
\label{sec:concept}
\begin{figure}[htbp]
    \centering
    \includegraphics[width=0.6\linewidth]{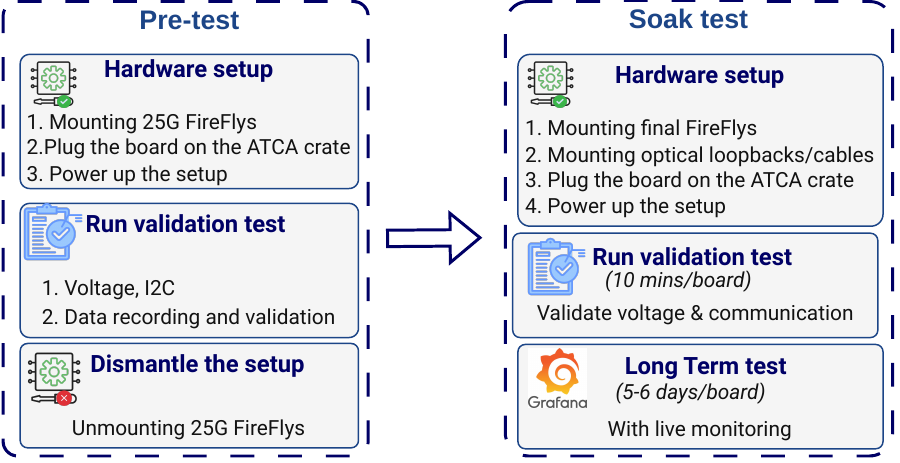}
    \caption{Overview of the two-phase OAT workflow.} 
    \label{fig:oat_workflow}
\end{figure}

The OAT is designed to be the first optical functional test 
applied to each fully assembled HGCAL Serenity-S1 card. It is 
performed in an ATCA test crate at room temperature and is 
structured into two sequential phases, as illustrated in 
Figure~\ref{fig:oat_workflow}.

\paragraph{Pre-test.} 
A rapid electrical pre-test is performed on every FireFly site using a dedicated set of 25\,Gbps FireFly modules. The pre-test aims to verify two conditions at each site:  I$^2$C communication with the FireFly module and the \text{supply rail voltage} for each inserted module. 

Board-level control is provided by the Serenity MAnagement SHell (SMASH)~\cite{smash}, which exposes peripheral 
components in the payload area via I$^2$C, PMBus, and a JTAG 
connection to the FPGA. 

\paragraph{Soak test.} 
The board is then equipped with its 
production FireFly modules, optical connections are made 
according to the board variant (listed in Section~\ref{sec:intro}). At this stage, a ``NullAlgo'' payload is deployed within the firmware (as described in Section~\ref{sec:impl_fw}), which routes the GTY transceivers directly to a PRBS31 pattern generator and checker. All optical links are then exercised continuously for up to five 
days.  Additionally, FireFly health parameters (temperature, voltages, optical receive power)
are logged at regular intervals. A thermal safeguard implemented in the OAT host software monitors FireFly temperatures and initiates a controlled shutdown of the FPGA and FireFly modules if any temperature exceeds a predefined threshold.

The required Bit Error Ratio (BER) target and the corresponding minimum test duration are determined using the confidence-level formulation for error-free transmission~\cite{Mitic_2012}. Assuming independent bit errors and a constant BER, the upper confidence bound for zero observed errors is
\begin{equation}
    \mathrm{BER} < \frac{-\ln(1-C)}{N_\mathrm{bits}},
\end{equation}
where $N_\mathrm{bits}$ is the total number of bits transmitted per link. Table~\ref{tab:ber} gives the limits for a five-day
soak test at each OAT link speed. At 95\% confidence, all speeds
reach $\mathrm{BER} < 10^{-15}$. At 99\% confidence, 10\,Gbps
links only just miss this value. 
For comparison, the five-second copper loopback of the
FAT at 25\,Gbps gives a limit of only $\approx\num{3.8e-11}$  (95\% C.L.)~\cite{Krause_2026}.

\begin{table}[ht]
\centering
\caption{BER upper limits for zero observed errors after a
five-day soak test, per link.}
\label{tab:ber}
\adjustbox{width=0.8\linewidth}{
\begin{tabular}{cccccc}
\hline
Link speed & $N_\mathrm{bits}$ & \multicolumn{2}{c}{BER upper limit} &
\multicolumn{2}{c}{Time to reach $10^{-15}$} \\
(Gbps) & (5 days) & 95\% C.L. & 99\% C.L. & 95\% C.L. & 99\% C.L. \\
\hline
10 & \num{4.3e15}  & \num{6.9e-16} & \num{1.1e-15} & 3.5 d  & 5.3 d \\
16 & \num{6.9e15}  & \num{4.3e-16} & \num{6.7e-16} & 2.2 d  & 3.3 d \\
25 & \num{1.1e16}  & \num{2.8e-16} & \num{4.3e-16} & 1.4 d  & 2.1 d \\
\hline
\end{tabular}}
\end{table}

A board passes only if: zero bit errors are observed on all links
(no PRBS lock loss); I$^2$C communication is confirmed at all
FireFly sites; the 3.75\,V rail is operational at all 25\,Gbps-capable
sites; and FireFly temperatures remain within the safe operating
range throughout. Any failure is documented in an auto-generated
PDF report and triggers rework before the board is released for integration.

\section{Implementation}
\label{sec:impl}
\subsection{Firmware}
\label{sec:impl_fw}

The OAT FPGA firmware is based on the Extensible Modular Processor (EMP) framework (EMP-FWK)~\cite{emp_fwk}, which separates a user-specific payload area from an infrastructure framework supporting multiple link protocols (CSP at 
16\,Gbps and 25\,Gbps,
as well as lpGBT) configurable per MGT quad. Since the infrastructure already provides BER Test (BERT) capabilities, no dedicated payload is needed for the OAT. While CSP16 and CSP25 natively expose the MGTs eyescan circuitry to IPBus~\cite{ipbus}, as required for BERT testing, this was not originally implemented for lpGBT links. This led to the development of a novel CSP10 PRBS-only protocol to also support BERT testing at 
10\,Gbps.
Additionally, mixed TX/RX protocol support per MGT quad was implemented to enable asymmetric speed testing required for the Trigger Stage-1 Serenity-S1 board validation. 

\subsection{Software} 
\label{sec:impl_sw} 
The OAT software consists of two complementary components: a link-validation system and a hardware-monitoring framework.

To accommodate the different variants of the HGCAL board, the link-validation software supports two test topologies: a loopback topology, used for DAQ and NZS boards, in which each transmit channel is connected to its corresponding receive channel, and a board-to-board topology, used for Stage-1 and Stage-2 boards, in which a shared test board provides the optical connections.
The topology and test parameters are specified in a YAML file.

Additionally, a thermal safeguard operates throughout the soak test. If any monitored FireFly temperature exceeds a predefined threshold, the FPGA and FireFly modules are powered off, and the test terminates with an error code.

\paragraph{Link validation.}
The link-validation software comprises two Python scripts. The host-side script deploys software and firmware packages via \texttt{rsync}, programs the FPGA, launches the on-board test script, and retrieves the results via SCP. The on-board script initializes the EMP controller and SMASH session, configures the transmitter (TX) and receiver (RX) MGTs for PRBS31 operation, and monitors PRBS lock status and FireFly health metrics. At the end of the test, bit-error results are evaluated and stored in json and csv files.

\paragraph{Hardware monitoring and visualization.}
To support long-term OAT runs, a complementary hardware-monitoring framework was developed for Serenity boards running HERD~\cite{HERD_Shepherd}. HERD collects board-level telemetry, including FPGA temperature and power metrics, within an on-board container and exposes the data to the external network. The monitoring interface is accessed remotely through the existing SSH tunnel, enabling persistent logging and real-time visualization with Grafana~\cite{Grafana}.

\begin{figure}[htbp]
    \centering
    \begin{minipage}[c]{0.4\textwidth}
        \centering
        \includegraphics[width=0.70\textwidth]{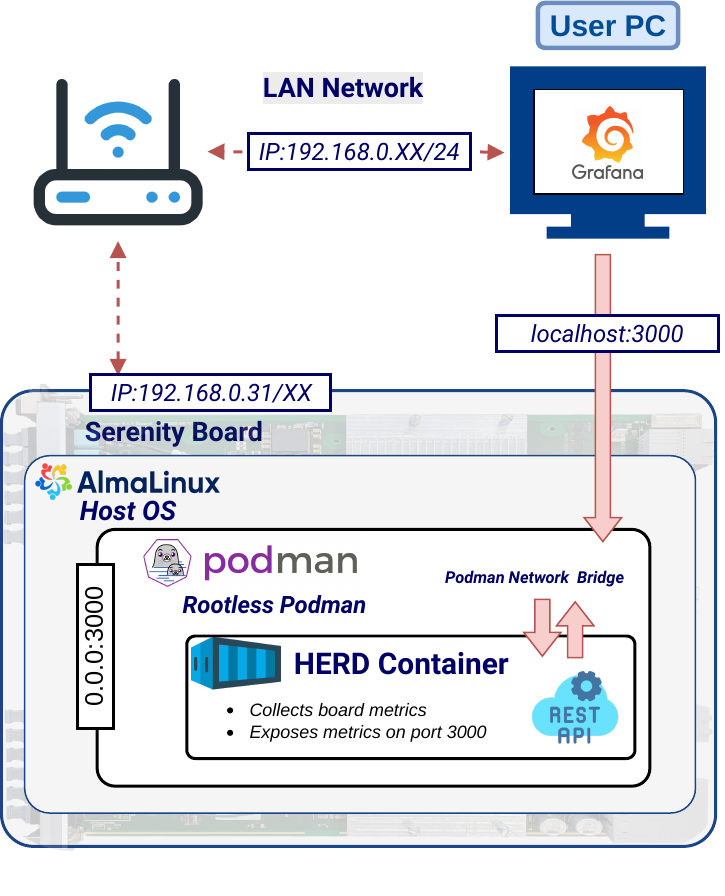}
        \label{fig:hw_oat_arch}
    \end{minipage}
    \hfill
    \begin{minipage}[c]{0.58\textwidth}
        \centering
        \includegraphics[width=\textwidth]{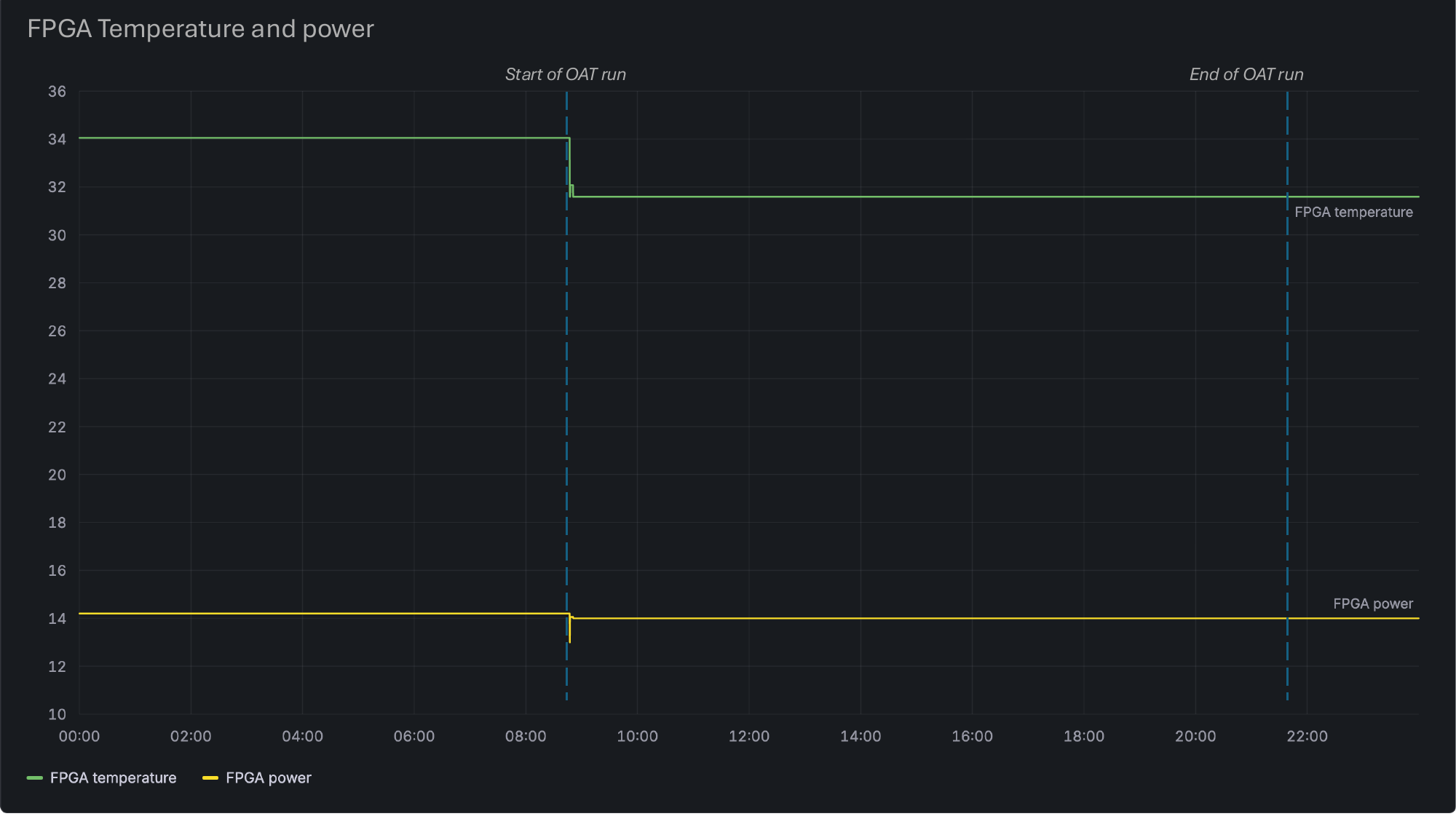}
        \label{fig:hw_oat_grafana}
    \end{minipage}
    \caption{HERD hardware monitoring for Serenity boards: (a) Telemetry collection and access
        architecture; (b) Grafana monitoring of FPGA temperature
        and power during 24-hour scan.}  
    \label{fig:hw_oat_herd}
\end{figure}

Figure~\ref{fig:hw_oat_herd} shows the monitoring architecture and an example of its application during a 24-hour scan using the OAT for a DAQ board.
The Grafana plot shows the evolution of FPGA temperature (green) and power (amber) over the scan, demonstrating the framework's ability to track board operating conditions throughout data acquisition.

\section{Hardware Setup and Commissioning Results}
\label{sec:hw_results}
The OAT was commissioned on the first HGCAL Serenity-S1 boards received at CERN, populated with FireFly modules. Two board flavours have been tested to date: DAQ and Trigger Stage-1.

The results presented here demonstrate initial functional operation of both test topologies; the full five-day soak-test duration and associated BER sensitivity have not yet been demonstrated.

\begin{figure}[htbp]
    \centering
    \begin{minipage}[c]{0.48\textwidth}
        \centering
        \includegraphics[width=\linewidth]{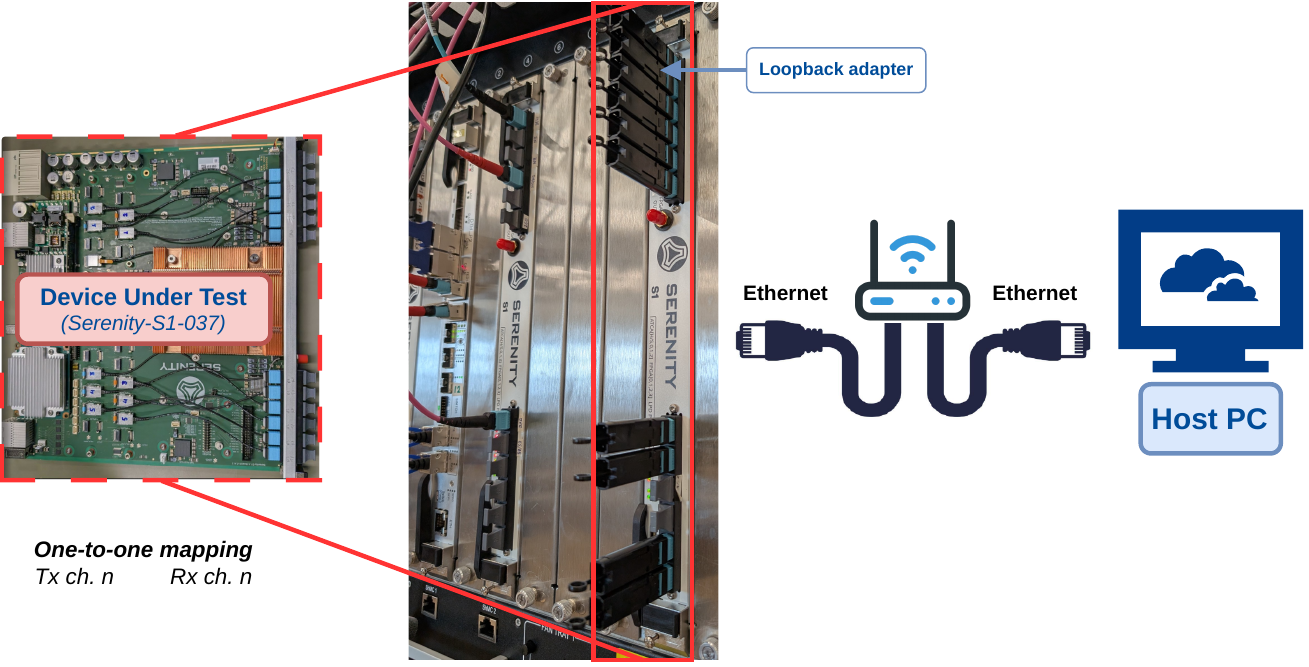}
    \end{minipage}\hfill
    \begin{minipage}[c]{0.48\textwidth}
        \centering
        \includegraphics[width=\linewidth]{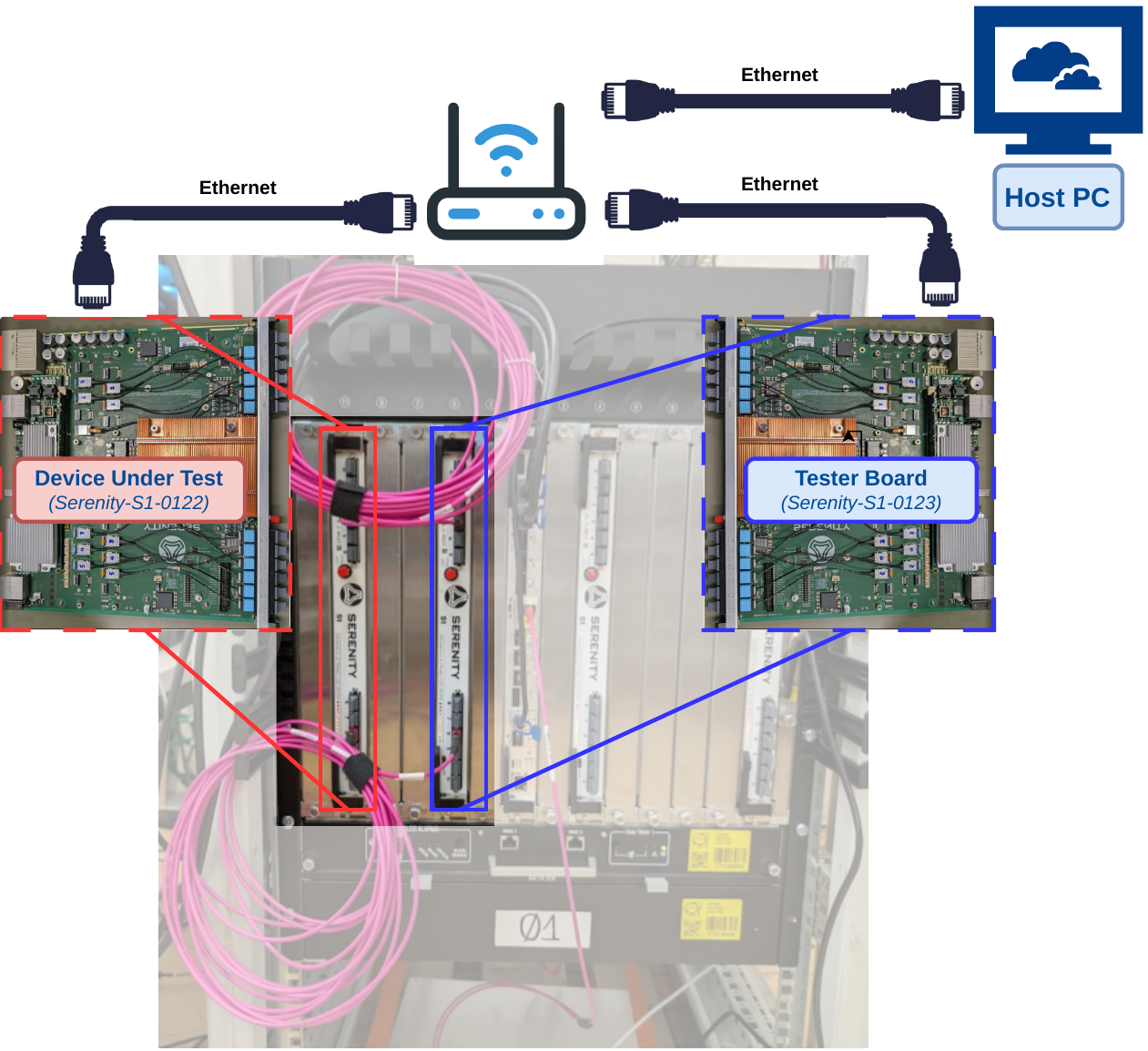}
    \end{minipage}
    \caption{Hardware setups for OAT optical-link validation.
    Left: DAQ board in the one-to-one optical loopback topology.
    Right: Trigger Stage-1 and tester board-to-board topology.}
    \label{fig:hw_oat}
\end{figure}

Figure~\ref{fig:hw_oat} shows the two test setups. The DAQ board is installed in an ATCA crate for optical loopback testing (left), while the Stage-1 boards are connected in a board-to-board topology (right).

In both setups, the device under test (DUT) and, for board-to-board testing, the complementary tester board, are controlled by a host PC over Ethernet through a network router. In the loopback topology, each FireFly transmit channel is connected to its corresponding receive channel using a loopback adapter, allowing optical links to be tested without a second board.\footnote{FireFly links intended to operate at 2.5\,Gbps in the final system are tested at 10\,Gbps.} In the board-to-board topology, the transmit channels of the DUT and tester board are connected to the corresponding receive channels on the other board, enabling bidirectional optical-link testing.

In the final OAT, one of five firmware variants will be loaded into the FPGAs on the Serenity-S1 based on the role of the board: \textit{DUT-DAQ}, \textit{DUT-Trigger Stage-1/2}, and \textit{Tester-Trigger Stage-1/2}.

\paragraph{DAQ board (April 2026).}
The electrical pre-test passed at all FireFly sites, confirming I$^2$C communication and 3.8\,V supply integrity. A 24-hour optical loopback soak test was then performed for commissioning. PRBS31 lock was maintained on all links, with no bit errors observed throughout the test. 

\begin{figure}[htbp]
    \centering
    \includegraphics[width=0.8\textwidth]{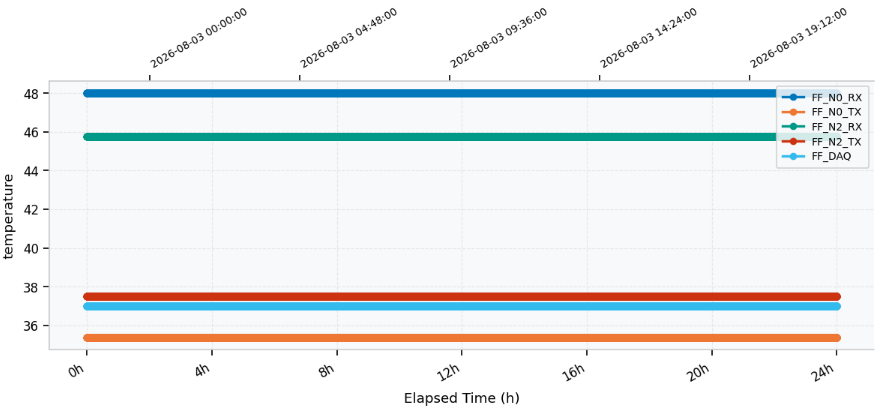}
    \caption{FireFly module temperatures during the optical loopback soak test.}
    \label{fig:firefly_temperature}
\end{figure}

FireFly health parameters were monitored throughout the test to check that the modules operated within their specified limits. Figure~\ref{fig:firefly_temperature} shows the temperatures recorded for several FireFly modules during the optical loopback soak test. 
The thermal guard was configured with a threshold of \SI{50}{\celsius}, following the guidance in the Samtec FireFly FAQ~\cite{samtec_firefly_faq}. 

\paragraph{Trigger Stage-1 board and tester (August 2026).}
The electrical pre-test passed on both boards. A five-minute board-to-board soak test confirmed PRBS lock on all links, with no bit errors observed. A full five-day soak test is planned as the next step.

These commissioning results demonstrate that both OAT topologies are operational and that the firmware supports the asymmetric link configuration of Stage-1 boards. A large-scale OAT campaign is planned to begin in 2027, covering all boards received by HGCAL as they are delivered.

HGCAL boards are expected to arrive in batches of approximately 20 boards per month. The OAT setup can test up to 20 boards in parallel, with an estimated processing time of 5–6 days per batch (approximately one to two weeks including setup and retesting). This allows each batch to be tested before the next arrives. The number of parallel test stations determines the required test equipment, including optical and electrical cables, and the associated cost.


\section{Conclusion and Outlook}
\label{sec:conclusion}
A fully automated OAT framework has been developed and
commissioned for HGCAL Serenity-S1 boards. It supports all four
board variants across loopback and board-to-board topologies, with
a novel CSP10 PRBS-only protocol to enable PRBS-based bit-error testing at 10 Gbps.
HERD and Grafana integration provide online monitoring for parallel
multi-board production runs. Commissioning results on DAQ and
Stage-1 boards confirm that the scheme is ready for large-scale
deployment, beginning in 2027. 

The remaining steps needed are finalising
the test-result database, procuring cables and tester boards, and
operator training.

\acknowledgments
H. A.~Krause acknowledges the support of the Doctoral School "Karlsruhe School of Elementary and Astroparticle Physics: Science and Technology" (KSETA). We also thank D.~Monk and T.~Williams for developing and adding the CSP10 PRBS-only protocol to the EMP-FWK.


\begin{thebibliography}{99}

\bibitem{cms_phase2_daq}
C. Amendola et al.,
\textit{An overview of the CMS High-Granularity Calorimeter},
JINST \textbf{v20} (2025) C01031,
doi: \url{https://doi.org/10.1088/1748-0221/20/01/C01031}.

\bibitem{Mehner_2025}
T. Mehner et al.,
\textit{Serenity-S1 -- A Versatile ATCA Processing Card for the CMS Phase-2 Upgrade},
TechRxiv (2025),
doi: \url{https://doi.org/10.36227/techrxiv.176162104.42530921/v1}.

\bibitem{Mehner_2024}
T. Mehner et al.,
\textit{Lessons learned while developing the Serenity-S1 ATCA card},
JINST \textbf{v19} (2024) C02018,
doi: \url{https://doi.org/10.1088/1748-0221/19/02/C02018}.

\bibitem{cms_phase2_tdr}
CMS Collaboration,
\textit{The Phase-2 Upgrade of the CMS Data Acquisition and High Level Trigger},
CERN-LHCC-2021-007, CMS-TDR-022 (2021),
\url{https://cds.cern.ch/record/2759072}.

\bibitem{cms_phase2_cms_trigger}
CMS Collaboration,
\textit{The Phase-2 Upgrade of the CMS Level-1 Trigger},
CERN-LHCC-2020-004, CMS-TDR-021 (2020),
\url{https://cds.cern.ch/record/2714892}.

\bibitem{Krause_2026}
H. Krause et al.,
\textit{Factory Acceptance Test for Serenity-S1, a CMS Phase-2 back-end card},
JINST \textbf{v21} (2026) C03035,
doi: \url{https://doi.org/10.1088/1748-0221/21/03/C03035}.

\bibitem{smash}
SERENITY Collaboration,
\textit{All Things SMASH} (2019),
\url{https://serenity.web.cern.ch/serenity/smash/index.html}.

\bibitem{Mitic_2012}
D. Mitić et al.,
\textit{Calculating the Required Number of Bits in the Function of Confidence Level and Error Probability Estimation},
Serbian Journal of Electrical Engineering \textbf{v9} (2012) no. 3, 361--375,
doi: \url{https://doi.org/10.2298/SJEE1203361M}.

\bibitem{emp_fwk}
K. Adamidis et al.,
\textit{EMP: a common infrastructure firmware framework for the CMS phase-2 upgrades},
JINST \textbf{v21} (2026) C02019,
doi: \url{https://doi.org/10.1088/1748-0221/21/02/C02019}.

\bibitem{ipbus}
C. Ghabrous Larrea et al.,
\textit{IPbus: a flexible Ethernet-based control system for xTCA hardware},
JINST \textbf{v10} (2015) C02019,
doi: \url{https://doi.org/10.1088/1748-0221/10/02/C02019}.

\bibitem{HERD_Shepherd}
CERN,
\textit{Shepherd Framework Documentation: FSM Runner Scripts},
\url{https://shepherd-fwk.docs.cern.ch/user-guide/fsm-runner-scripts/index.html}.

\bibitem{Grafana}
Grafana Labs,
\textit{Grafana Documentation},
\url{https://grafana.com/docs/grafana/latest/}.

\bibitem{samtec_firefly_faq}
Samtec,
\textit{FireFly FAQ},
\url{https://web.pa.msu.edu/hep/atlas/l1calo/hub/hardware/components/optical/samtec_firefly/samtec_firefly_faq.pdf}.

\end{thebibliography}
\end{document}